\documentclass[%
 reprint,
superscriptaddress,
 amsmath,amssymb,
 aps,
]{revtex4-1}

\usepackage{graphicx}
\usepackage{dcolumn}
\usepackage{bm}
\usepackage[ansinew]{inputenc}
\usepackage[dvipsnames]{xcolor}

\begin{document}

\preprint{APS/123-QED}

\title{Coupling a single trapped atom to a whispering-gallery-mode microresonator}

\author{Elisa Will}
 \affiliation{Vienna Center for Quantum Science and Technology, Atominstitut, TU Wien, Vienna, Austria.}
\author{Luke Masters}
 \affiliation{Vienna Center for Quantum Science and Technology, Atominstitut, TU Wien, Vienna, Austria.}
   \affiliation{Department of Physics, Humboldt-Universit\"at zu Berlin, Germany.}
\author{Arno Rauschenbeutel}
 \affiliation{Vienna Center for Quantum Science and Technology, Atominstitut, TU Wien, Vienna, Austria.}
 \affiliation{Department of Physics, Humboldt-Universit\"at zu Berlin, Germany.}
\author{Michael Scheucher}
 \affiliation{Vienna Center for Quantum Science and Technology, Atominstitut, TU Wien, Vienna, Austria.}
\author{Jürgen Volz}
 \affiliation{Vienna Center for Quantum Science and Technology, Atominstitut, TU Wien, Vienna, Austria.}
 \affiliation{Department of Physics, Humboldt-Universit\"at zu Berlin, Germany.}


\begin{abstract}
We demonstrate trapping of a single $^{85}$Rb atom at a distance of 200~nm from the surface of a whispering-gallery-mode bottle microresonator. The atom is trapped in an optical potential, which is created by retroreflecting a red-detuned focused laser beam from the resonator surface. We counteract the trap-induced light shift of the atomic transition frequency by superposing a second laser beam with suitably chosen power and detuning. This allows us to observe a vacuum Rabi-splitting in the excitation spectrum of the coupled atom-resonator system. This first demonstration of stable and controlled interaction of a single atom with a whispering-gallery-mode in the strong coupling regime opens up the route towards the implementation of quantum protocols and applications that harvest the chiral atom-light coupling present in this class of resonators.
\end{abstract}


\pacs{Valid PACS appear here}
                              
\maketitle
In free space, the interaction between single atoms and single photons is weak. However, by strongly confining the photons inside a microresonator with high quality factor, $Q$, and by coupling the atoms to the resonator mode, atom-light interaction can be significantly enhanced. This approach lies at the heart of cavity quantum electrodynamics (CQED) \cite{thompson1992observation,brune1996quantum,kuhn2002deterministic}. Over the past decades, ultra-high finesse Fabry-P\'erot microresonators have been instrumental in advancing this field. In these traditional resonators, single atoms have been trapped inside the resonator mode \cite{ye1999trapping,pinkse2000trapping}, which has lead to many ground-breaking experiments \cite{reiserer2015cavity}. 

More recently, also other resonator types have successfully been employed in single-atom CQED, such as optical fiber-based Fabry-P\'erot cavities \cite{volz2011measurement,steiner2013single,gallego2018strong,brekenfeld2020quantum}, photonic crystal cavities \cite{thompson2013coupling,goban2015superradiance}, optical nanofiber-based cavities \cite{kato2015strong,nayak2019real}, and whispering-gallery-mode (WGM) microresonators \cite{aoki2006observation,junge2013strong,shomroni2014all}. Out of those, WGM resonators distinguish themselves by offering chiral, i.e.~propagation-direction-dependent, light-matter interaction \cite{junge2013strong,lodahl2017chiral}, which enables novel protocols and functionalities for processing light on the quantum level \cite{o2013fiber,shomroni2014all,scheucher2016quantum,bechler2018passive}. However, so far, only free-falling atoms have been coupled to WGM resonators, resulting in a limited interaction time, a position-dependent coupling strength between atom and light which reduces process fidelities, and only a probabilistic operation. In this respect, CQED with WGM resonators now stands at the point where CQED with Fabry-P\'erot resonators was two decades ago, and gaining the ability to trap atoms inside the resonator field would be a crucial step forward.

\begin{figure}[h!]
	\centering
	\includegraphics[width=\columnwidth]{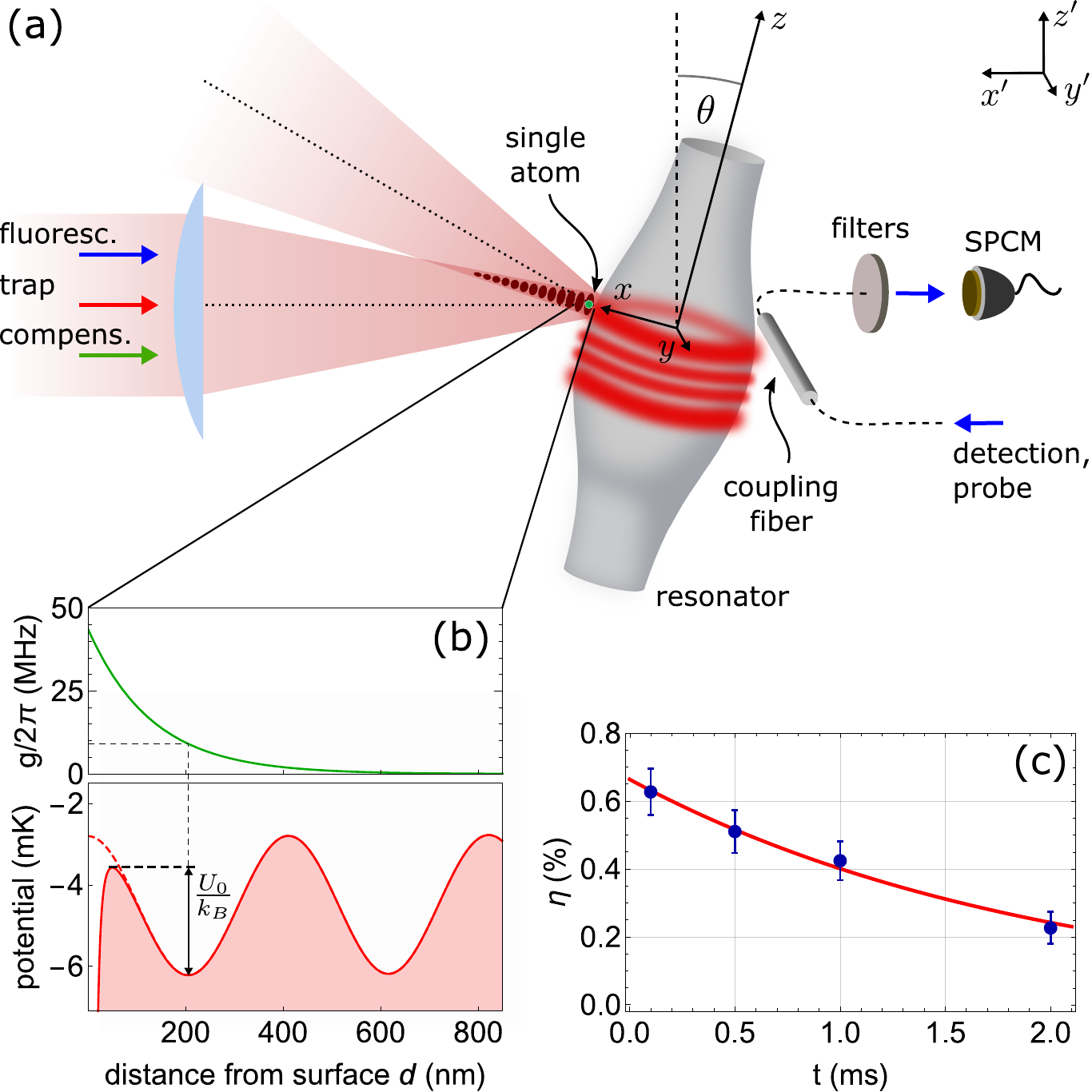}
	\caption{(a) Schematic of the experimental setup for detecting and trapping a single $^{85}$Rb atom close to a bottle microresonator in a standing-wave optical dipole trap. The optics for the dipole trap is also used to focus a laser beam onto the atom for fluorescence detection, as well as a laser beam for light shift compensation (see main text). The resonator-centered coordinate system $\left( x,y,z\right) $ and a laboratory-centered coordinate system $\left( x^\prime,y^\prime,z^\prime \right) $ are defined as indicated. (b) Trap potential (bottom panel) and expected coupling strength, $g$ (top panel), along the $x$-direction. (c) Measurement of the probability for finding a given detected single atom in the trap as a function of the waiting time $t$. The red line is an exponential fit, yielding a trap lifetime of $\tau \approx 2$ ms.}
	\label{fig:experimental_setup}
\end{figure} 

Here, we demonstrate strong coupling of a single trapped rubidium atom to a WGM bottle microresonator with ultrahigh quality factor, $Q$ \cite{pollinger2009ultrahigh}. We achieve this by overcoming two crucial experimental challenges: First, efficient coupling to the evanescent field of the WGM requires the atom to be trapped at a small distance from the resonator surface where van der Waals- and Casimir-Polder-forces have to be counteracted. To achieve this, we employ a deep standing-wave optical dipole trap, created by retroreflecting a focused trapping light field from the resonator surface \cite{thompson2013coupling}. Secondly, the intense light field of the trap induces a large position-dependent light-shift, which detunes the atomic resonance from the resonator mode. We effectively counteract the light shift of the atomic transition by means of a second, detuned compensation light field that shifts the excited state back into resonance \cite{hilton2019dual}. As a result, we observe a resonant vacuum Rabi-splitting in the excitation spectrum of the trapped atom-resonator system, indicating strong coupling.

The core elements of the experimental setup are shown in Fig.~\ref{fig:experimental_setup}(a). 
The WGM bottle microresonator features a quality factor of $Q \approx 5 \times 10^7$ and is stabilized to the resonance of the unperturbed $\left( 5 S_{1/2}, F=3\right)  \rightarrow \left( 5 P_{3/2}, F'=4\right)$ transition of $^{85}$Rb with angular frequency $\omega_0$. In order to couple light into and out of the resonator, it is interfaced with a tapered fiber coupler. The system is set to critical coupling such that, in the absence of a coupled atom, the transmission of resonant light through the coupling fiber vanishes. This occurs when the fiber-resonator coupling rate $\kappa_{\text{ext}}$ equals the intrinsic resonator field decay rate $\kappa_0$. The total resonator field decay rate is then given by $\kappa=\kappa_0+\kappa_{\text{ext}} \approx 2 \pi \times 10$~MHz, see below.

The optical dipole trap is created by retroreflecting a focused laser beam (waist radius: $w_\text{trap} = 3.5 \pm 0.3$~$\mu$m) from the bottle resonator surface. The wavelength of the trapping light ($\lambda_\text{trap}=783.68$~nm) is red-detuned with respect to the $5 S_{1/2} \rightarrow 5 P_{3/2}$ transition, see Fig. 2.
The interference between the incident beam and its reflection forms a partially modulated standing-wave pattern along the $x$-axis, orthogonal to the resonator axis ($z$-axis). The incidence angle $\theta \approx$ 17° with respect to the $x$-axis prevents unwanted interference between the incident beam and its reflection from the back-surface of the resonator. Figure \ref{fig:experimental_setup}(b) shows the trap potential and the expected atom-light coupling strength as a function of the distance, $d$, of the atom from the resonator surface in $x$-direction. We trap a single atom in the potential minimum closest to the resonator, which is located at $d_\text{trap}=\lambda_\text{trap}/\left( 4 \; \text{cos} \theta \right) \approx$~205~nm. 
For this position, we expect an atom-light coupling strength of $g \approx 2 \pi \times 10$~MHz which puts the system at the onset of strong coupling, i.e., $g > (\kappa,\gamma)$, where $\gamma= 2 \pi \times 3$~MHz is the atomic dipole decay rate. For the parameters that are used in the remainder of the manuscript, the axial and transverse trap frequencies are $ \left\lbrace  \omega_x,\omega_{y,z} \right\rbrace  \approx 2 \pi \times \left\lbrace 1 \, \text{MHz}, 60 \, \text{kHz} \right\rbrace $, respectively.

The trap is loaded from a 1-cm diameter cloud of about $5 \times 10^7$ laser-cooled $^{85}$Rb atoms that is delivered to the resonator by an atomic fountain and has a temperature of about 7~$\mu$K. In order to detect the presence of a single atom in the resonator mode in real time, we send resonant detection light through the coupling fiber and monitor the transmitted power using a single-photon counting module (SPCM). The interaction of an atom with the resonator mode leads to a transmission increase, to which we react using a field programmable gate array (FPGA)-based detection and control system \cite{junge2013strong}. Upon detection of an atom, the detection light is switched off using an electro-optical modulator, and the dipole trap is switched on using an acousto-optical modulator. The overall delay between detection and trapping is about 250~ns. As this duration is much shorter than the average transit time of an atom through the evanescent field of the resonator mode, it allows us to catch an atom which was detected in the vicinity of the trapping potential.

In order to verify that the trap loading succeeded, we use a fluorescence detection scheme: we launch a 20 $\mu$s-long detection pulse through the trap optics onto the atom. This light pulse has a center frequency of $\omega_0$ and a peak intensity of about 100 $I_\text{sat}$, where $I_\text{sat}$ is the saturation intensity of the $\left(5 S_{1/2}, F=3\right) \rightarrow \left(5 P_{3/2}, F^\prime=4\right)$ transition. A certain fraction of the fluorescence photons are scattered into the resonator mode and detected with the SPCM, see Fig.~\ref{fig:experimental_setup}. This allows us to detect the presence of an atom in the dipole trap in spite of the large trap-induced detuning between the atomic resonance and the resonator. By varying the timing of the light pulse, we measure the probability for finding a single atom, which was initially detected via increased resonator transmission, at a waiting time, $t$, after switching on the trap, see Fig.~\ref{fig:experimental_setup}(c). Fitting an exponential function (red solid line) to the measured probabilities, $\eta(t)=\eta_0 \exp(-t/\tau)$, yields a probability for trapping a given detected single atom of $\eta_0 \approx 0.7$~\% and a trap lifetime of $\tau \approx 2$ ms. This trapping probability is compatible with our expectation considering the finite overlap of the trap volume with the resonator mode, the initial kinetic energy of the atom and the time delay between atom detection and switching on the trap.

\begin{figure}[tb]
	\centering
	\includegraphics[width=\columnwidth]{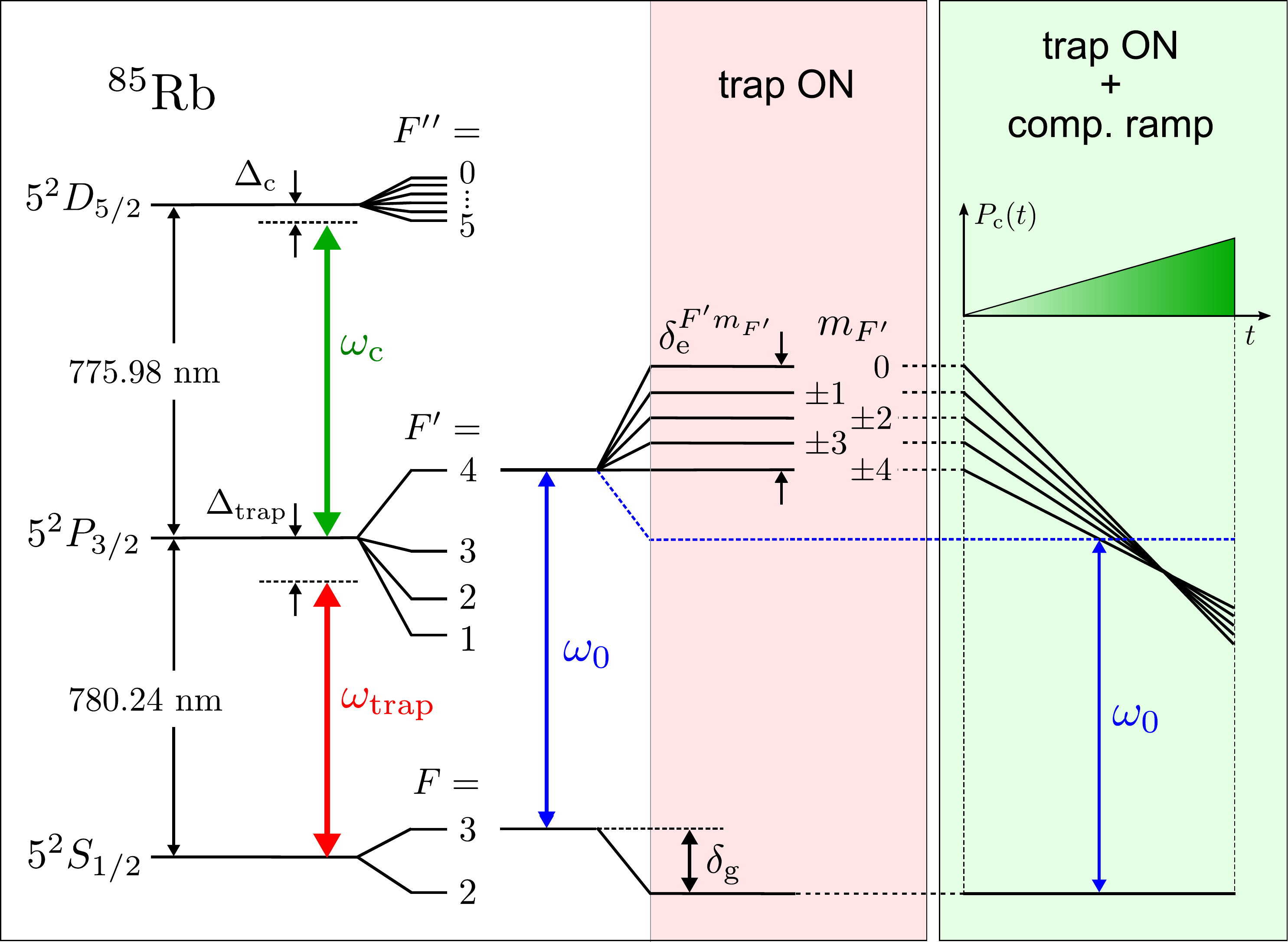}
	\caption{Principle of the light shift compensation. The left box shows the energy level structure of $^{85}$Rb relevant for the compensation scheme and illustrates the trap-induced light shifts of the $\left(5 S_{1/2}, F=3\right) \rightarrow \left(5 P_{3/2}, F^\prime=4\right)$ atomic transition, for a fixed power, $P_\text{trap}$. The right box qualitatively shows how the $F'=4$-Zeeman levels are tuned across resonance, when the compensation laser power, $P_\text{c}$, is increased in the presence of the trap light. A quantitative treatment of the compensation mechanism is shown in Fig. \ref{fig:LSC_power_scan}(a).}
	\label{fig:Level_scheme}
\end{figure}

The energy of an atom in the trap is predominantly determined by its kinetic energy and its position with respect to the trap center at the moment the dipole trap switches on. The kinetic energy is dominated by the free-fall of the atom and corresponds to $E/k_B\approx0.5$~mK. Consequently, trapping the atom requires an optical potential with significant depth. We use a dipole trap beam, which is linearly polarized and has a power of $P_\text{trap} = 19$~mW. When the trap field is polarized along the $y$-axis ($z^\prime$-axis), the trap depth amounts to $U_0/k_\text{B} \approx 2.6$~mK $\left( \approx 1.9 \;\text{mK} \right)$, due to the polarization-dependent reflection coefficient.

The trapping light field induces a scalar light shift, $\delta_\text{g}$, of the $5S_{1/2}$ ground state and a scalar as well as tensor light shifts, $\delta_\text{e}^{F^\prime m_{F^\prime}}$, of the $5P_{3/2}$ excited state, see Fig.~\ref{fig:Level_scheme}. The resulting Zeeman state-dependent shifts of the $5S_{1/2} \rightarrow 5P_{3/2}$ transition frequencies reach up to about 250~MHz. 

In order to efficiently interface the trapped atom with the resonator in the presence of this light shift, we have to compensate for the latter. For this purpose, we expose the atom to an additional light field -- the so-called compensation field. Its polarization is aligned with the linear polarization of the trap field and its frequency is red-detuned by $\Delta_\text{c}=2\pi\times927$~MHz with respect to the $5 P_{3/2}\rightarrow5 D_{5/2}$ transition. The compensation field induces a scalar and  tensor light shift on the $5 P_{3/2}$ excited state while its effect on the ground state is negligible. The compensation laser is offset-locked to the $5P_{3/2} \rightarrow 5D_{5/2}$ transition using absorption spectroscopy \cite{kalatskiy2017frequency}, see supplementary material. 

For optimal compensation of the trap-induced light shift of the $5S_{1/2} \rightarrow 5P_{3/2}$ transition, the relative intensity distribution of the compensation field should match as much as possible that of the trapping field. This is facilitated by the fact that the two fields have similar wavelengths ($\lambda_\text{trap}=~783.68$~nm and $\lambda_\text{c}=775.98$~nm). In our setup, we simultaneously send them through the same fiber-coupled focusing optics, such that the positions of their foci, their waist radii, and their standing-wave patterns near the resonator closely match upon reflection off the resonator. This allows us to cancel the trap-induced scalar light shift of the transition in the entire trapping volume, when the power of the compensation field, $P_\text{c}$, is adequately chosen.

\begin{figure}[tb]
	\centering
	\includegraphics[width=7.7cm]{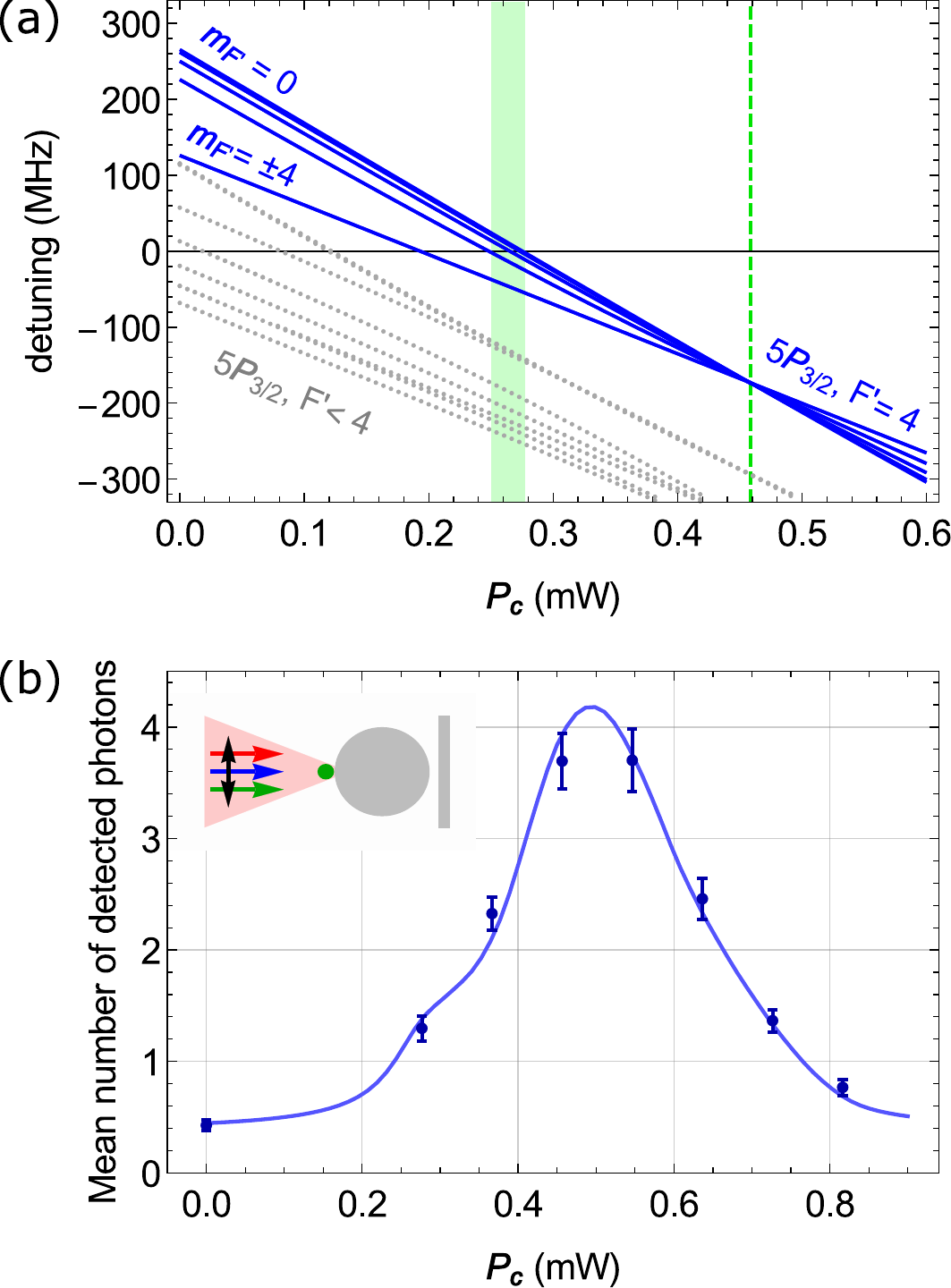}
	\caption{(a) Calculation of the detunings of the two-color light shifted frequencies of the transitions between the $ 5S_{1/2},F=3 $ ground state and the $5P_{3/2}$ excited state manifold with respect to $\omega_0$, as a function of the compensation laser power,~$P_\text{c}$, for $y$-polarized light. For $z^\prime$-polarized light, the light shifts are about $10$~\% smaller. The calculation is performed for the trap and compensation laser parameters given in the text and for an atom located at the trap center. The gray dotted lines are the detunings for the lower lying $F^\prime$-levels. (b) Measurement of the mean number of detected fluorescence photons in the presence of the trap and compensation lasers, as a function of the compensation laser power,~$P_\text{c}$.}
	\label{fig:LSC_power_scan}
\end{figure}

Figure \ref{fig:LSC_power_scan}(a) shows the calculated detunings of the two-color light shifted frequencies of the transitions between the $ 5S_{1/2},F=3 $ ground state and the $5P_{3/2}$ excited state manifold as a function of $P_\text{c}$, see supplementary material. Here, we assume that the atom is located in the trap center, and we take the frequency of the $F=3 \rightarrow F^\prime=4$ cycling transition in the absence of light shifts, $\omega_0$, as the reference. With increasing power of the compensation field, both the scalar and tensor light shifts of the transition frequencies decrease. 

In the power range from $P_\text{c} = 0.250$~mW to $0.276$~mW, the detuning of the transition frequencies to the ($F^\prime=4,|m_{F^\prime}| \leq 3$) excited states sequentially cross zero, see green shaded area. However, this does not imply that light shifts are vanishing, but that the light shifts of the ground and excited states are equal, see Fig. \ref{fig:Level_scheme}. The dashed green line indicates a second interesting setting. There, the transition frequencies to the Zeeman levels of each individual $F^\prime$-manifold intersect for the same value of $P_\text{c}$. In other words, the tensor light shift of the $5P_{3/2}$ excited state vanishes at this power and one approximately recovers the unperturbed excited state hyperfine structure, see supplementary material.

\begin{figure}[tb]
	\centering
	\includegraphics[width=7.5cm]{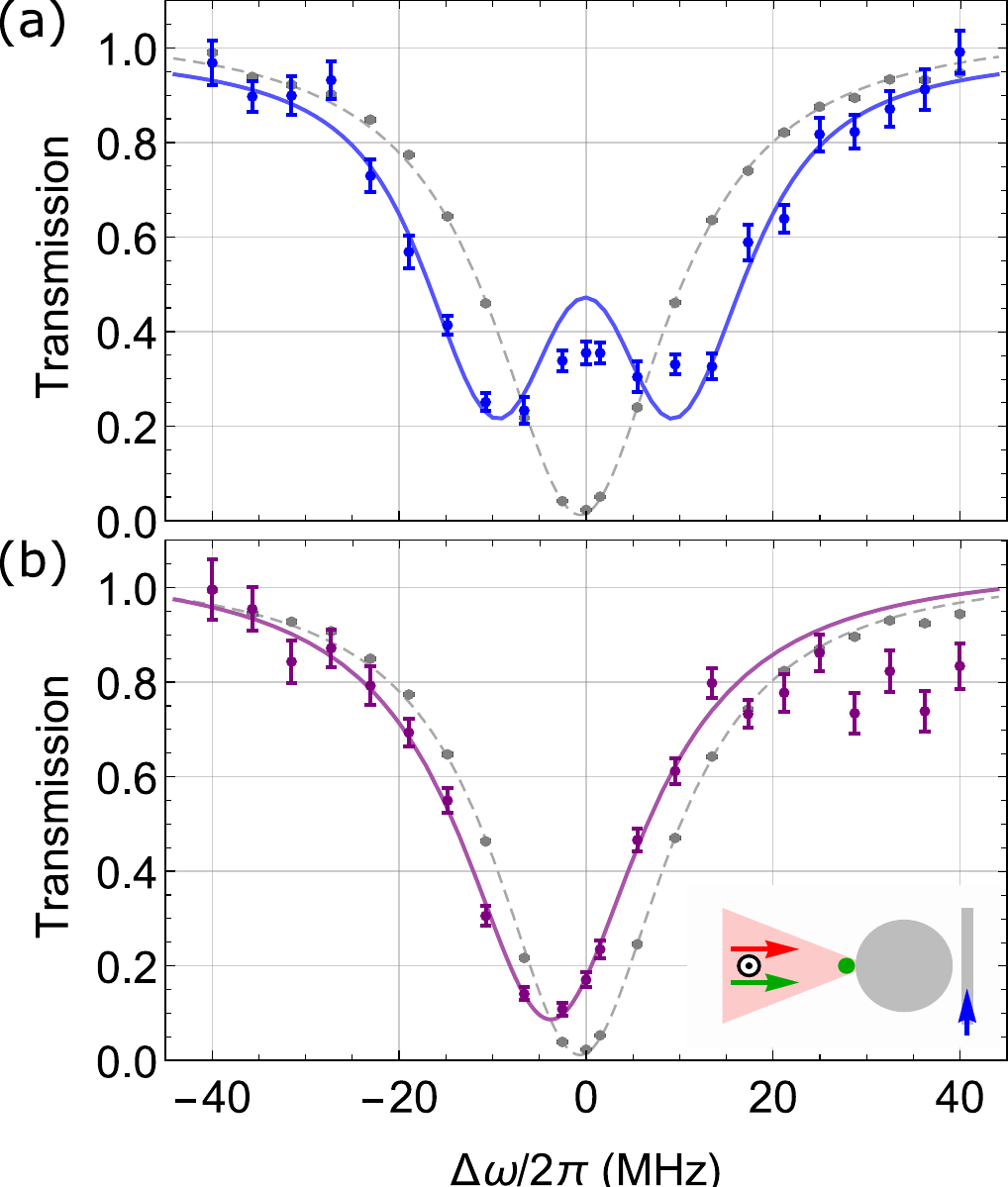}
	\caption{Normalized transmission spectra of the atom-resonator system (a) in presence ($P_\text{c} = 330 \pm 30$ $\mu$W) and (b) absence of the compensation laser. Each plot shows the average spectrum of single trapped atoms (blue and purple, respectively), as well as the empty resonator spectrum when no atom was trapped (gray). The error bars of the gray data points are smaller than the point size. The solid line in (a) is a theoretical prediction (see details in the text). The solid line in (b) is a fit of a Lorentzian function to the data, excluding the six right-most data points, as a guide to the eye. The dashed lines are fitted Lorentzians with a HWHM of about 11~MHz. The inset in (b) illustrates the propagation directions of the light fields, and the polarization direction of the trap and compensation fields.}
	\label{fig:Rabi_spectrum}
\end{figure}

In order to experimentally find the point of optimal compensation we measure the fluorescence of a single trapped atom emitted into the resonator mode, as a function of the compensation laser power. For this purpose, we send probe light resonant with the unperturbed $F=3 \rightarrow F^\prime=4$ transition through the trap optics for 100~$\mu$s and measure the intra-resonator power via the coupling fiber. In this measurement, the probe laser intensity was about 2 $I_\text{sat}$ at the center of the trap. 
Figure \ref{fig:LSC_power_scan}(b) shows the mean number of fluorescence photons detected via the coupling fiber. The blue solid line is a theoretical fit where we calculate fluorescence spectra for transitions between all Zeeman levels of the $F=3$ and $F^\prime=4$ manifolds including the position-dependent light shifts from the trap and the compensation laser fields, see supplementary material. Moreover, we expect that, because of chromatic aberrations of our trap optics the foci of the two beams are displaced with respect to each other along the beam axis. This results in a mismatch of the radii of the trapping and compensation laser beams at the position of the atom. We average the calculated spectra over the position distribution of the atoms in the trap, which we estimate from an independent temperature measurement, see supplementary material. We fit this average spectrum to the data using as fit parameters the beam radius mismatch, the maximum mean number of detected photons and the polarization ellipticity introduced from birefringence of the optical components. This fit yields a ratio of the beam radii of $w_\text{c}/w_\text{trap}=1.5$ and a reduced overlap with linear polarization of about 96~\%. Both values are reasonable for our experimental setup and the agreement between the theoretical fit and the experimental data is very good. The maximum scattering rate is found for a compensation laser power of $P_\text{c}^\text{max}= 500 \pm 50$~$\mu$W.

We now measure spectra of the coupled atom-resonator system for different compensation laser powers, $P_\text{c}$. Here, we align the linear polarizations of the dipole trap and of the compensation light along the $z'$-axis. With respect to this axis, the transverse-magnetic WGMs are approximately $\sigma^+$- or $\sigma^-$-polarized and, thus, couple to the $F=3, m_F=\pm 3 \rightarrow F^{\prime}=4, m_{F^{\prime}}=\pm 4$ cycling transition. 
 
We launch probe light through the coupling fiber and vary its detuning $\Delta \omega= \omega_\text{probe}-\omega_0$. For each value of $\Delta \omega$, we record the transmission, which is averaged over $400$~$\mu$s. Figure \ref{fig:Rabi_spectrum}(a) shows the most symmetric spectrum, which we obtained for $P_\text{c}=330 \pm 30$ $\mu$W. It exhibits a vacuum Rabi-splitting corresponding to an atom-photon coupling strength of about $g=2 \pi \times 10$~MHz, in agreement with our expectations, see Fig. \ref{fig:experimental_setup}(b). This demonstrates, for the first time, strong coupling of a single trapped atom to a WGM resonator. We note that, in addition to compensating the mean light shift of the atomic transition, the light shift compensation also reduces the substantial broadening of the atomic transition frequency and the concurrent blurring of the Rabi splitting due to the atomic motion in the trap. For comparison, Fig. \ref{fig:Rabi_spectrum}(b) shows the transmission spectrum in the absence of light shift compensation. In this case, due to the large atom-resonator detunings, we only observe a small dispersive shift of the resonator resonance. The spectra for the other values of $P_\text{c}$ are shown in the supplementary material.

The solid line in Fig. \ref{fig:Rabi_spectrum}(a) is the theoretical prediction of the vacuum Rabi spectrum for a compensation laser power of $P_\text{c}^\text{th}=400$~$\mu$W, where the spectrum turns out to be most symmetric. It was obtained by averaging vacuum Rabi spectra over different atom-photon coupling strengths and light shifts for the same position distribution and beam radius mismatch as the ones used for analysing Fig. \ref{fig:LSC_power_scan}(b). The resulting average coupling strength is $g^\text{th}/ 2 \pi = 9.3$~MHz with a standard deviation of $\sigma^\text{th}_g / 2 \pi=6.1 $~MHz. We note that, even though the experimentally determined value for $P_\text{c}^\text{max}$ is $\sim 25$~\% larger, a symmetric vacuum Rabi spectrum for $P_\text{c}^\text{th}$ around 400~$\mu$W is indeed expected when consulting Fig. \ref{fig:LSC_power_scan} (a). In measuring $P_\text{c}^\text{max}$, transitions between all magnetic sublevels of the $F=3$ and $F^\prime=4$ hyperfine manifolds contribute to the signal. For the measurement of the Rabi spectrum, however, we predominantly drive the $F=3, m_F= 3 \rightarrow F^{\prime}=4, m_{F^{\prime}}= 4$ cycling transition. As shown in panel (a), the latter features a smaller light shift, which is already compensated for a laser power of $\sim 0.75 P_\text{c}^\text{max}$. Still, there is a 17~\% discrepancy between the compensation laser powers for which we obtain a symmetric vacuum Rabi spectrum in the experiment ($P_\text{c}=330 \pm 30$~$\mu$W) and in the theoretical model ($P_\text{c}^\text{th}=400$~$\mu$W). We mostly attribute this to the $\pm 10$~\% error in determining $P_\text{c}^\text{max}$ from Fig.~\ref{fig:LSC_power_scan}(b). This error impacts our estimation of the ratio of the compensation and trap laser beam radii, $w_\text{c}/w_\text{trap}$, which enters as a parameter in our theoretical model. 

In summary, we optically trapped a single atom at a distance of about 200~nm from the surface of a WGM microresonator and compensated the position-dependent trap-induced light shift of the atomic transition. This allowed us to demonstrate stable and controlled interaction of a single atom with a whispering-gallery-mode in the strong coupling regime, thereby overcoming a long-standing challenge of optical cavity quantum electrodynamics. The demonstrated method can also be applied to WGM and ring resonators on integrated optical chips. This lays the pathway towards realizing more complex quantum-controlled photonic circuits, which may in particular profit from the chiral nature of atom-light coupling in this setting. 

We thank Adèle Hilico for her contributions in the early stages of the experiment. This work was financially supported by the Austrian Science Fund (FWF; SFB FoQuS Project No. F 4017 and DK CoQuS Project No. W 1210-N16), the Alexander von Humboldt-Foundation (Alexander von Humboldt Professorship). \\

\bibliography{citation}
\end{document}



\title{Supplementary Material: Coupling a single trapped atom to a whispering-gallery-mode microresonator}

\author{Elisa Will}
\affiliation{Vienna Center for Quantum Science and Technology, Atominstitut, TU Wien, Vienna, Austria.}
\author{Luke Masters}
\affiliation{Vienna Center for Quantum Science and Technology, Atominstitut, TU Wien, Vienna, Austria.}
\affiliation{Department of Physics, Humboldt-Universit\"at zu Berlin, Germany.}
\author{Arno Rauschenbeutel}
\affiliation{Vienna Center for Quantum Science and Technology, Atominstitut, TU Wien, Vienna, Austria.}
\affiliation{Department of Physics, Humboldt-Universit\"at zu Berlin, Germany.}
\author{Michael Scheucher}
\affiliation{Vienna Center for Quantum Science and Technology, Atominstitut, TU Wien, Vienna, Austria.}
\author{Jürgen Volz}
\affiliation{Vienna Center for Quantum Science and Technology, Atominstitut, TU Wien, Vienna, Austria.}
\affiliation{Department of Physics, Humboldt-Universit\"at zu Berlin, Germany.}


\maketitle

\subsection{Dipole trap optics} 
The optics assembly for the dipole trap beam is mounted outside the vacuum chamber in front of a viewport, which has an anti-reflection coating on both optical surfaces with $<0.1$~\% reflectance at 780~nm. The trap optics consists of two identical $75$ mm-diameter aspheric lenses with a focal length of $f=150$~mm (Thorlabs, AL75150-B) that focus the trap beam originating from a single-mode fiber (Thorlabs, 780HP; NA=0.13) onto the bottle resonator surface. The resonator is located inside the vacuum chamber at a distance of about 120~mm to the viewport. Given this geometry, we obtain a beam waist of $w \approx 3.5 $~$\mu$m. All three external light fields ($\lambda_\text{trap}$, $\lambda_\text{c}$ and $\lambda_0$), used in the presented measurements, originate from the same fiber and have the same polarization direction, which can be turned via a $\lambda/2$-waveplate between the two lenses.

\subsection{Positioning the trap focus on the resonator}

In order to obtain the desired trap potential and to yield an optimal trapping efficiency, the focus position of the trap beam has to be carefully adjusted with respect to the resonator in all three directions. For this purpose, the trap optics is mounted on a 3-axis translation stage (Thorlabs, NanoMax 300). The translation stage is equipped in the different axes with manual micrometer screw ($x^\prime$), stepper motor actuator ($y^\prime$) and piezo-electric actuator as well as manual micrometer screw ($z^\prime$). 
 
In order to position the beam waist on the resonator surface, we use the following procedure. We send a small amount of trap laser power ($\approx 50$ $\mu$W) through the trap optics. By measuring the number of photons coupled into the nanofiber with the SPCM, we probe the transverse intensity profile of the beam. Moving the position of the trap focus in $x^{\prime}$-direction then allows us to locate and position the beam waist. When the latter is at the position of the nanofiber, we translate the beam in axial direction by the diameter of the resonator ($\approx 36$~$\mu$m) to achieve the desired position. This alignment is not very critical, considering the Rayleigh length of about 50 $\mu$m of the trap beam.
 
The optimal position of the beam focus in $z^{\prime}$-direction along the resonator was identified by an independent measurement of the trapping efficiency as a function of the $z^{\prime}$-position. In $y^\prime$-direction the trap focus should be centered on the resonator $\left(y^\prime=0 \right) $. To facilitate the adjustment of the trap focus to the desired  $\left\lbrace y^\prime,z^\prime \right\rbrace $-position we create a 2D-map of the resonator and the nanofiber by measuring the position-dependent amount of light scattered into the nanofiber. For this purpose, we again send a weak trapping light through the trap optics and scan the beam focus across the resonator-fiber system in an area of $\Delta z^\prime \times \Delta y^\prime$ = 20~$\mu$m $\times$ 140~$\mu$m. In particular, this procedure makes the resonator edges and the nanofiber visible, providing a reference that enables reproducible transverse beam positioning with the required precision. 

\subsection{Delivery of atoms to the resonator \\and trap loading}
\label{sec:Exp_loading}

The experimental cycle, which takes about 1.7~s in total, starts with loading $^{85}$Rb atoms into a magneto-optical trap (MOT) and laser-cooling them for about 1~s to a temperature $\approx 7$ $\mu$K. The cold cloud is then launched to the bottle resonator by an atomic fountain and spends about 80~ms in close vicinity to the resonator. During this ''detection window'' we detect about 10 atoms strongly coupled to the resonator mode in real-time as described in the main part of the manuscript. 

After many repetitions of the experimental cycle, we observe a double-peak signal when measuring the atom detection probability as a function of time: first, when the atoms pass the resonator on their way upwards and a second time when atoms are free-falling after the turning point of the ballistic fountain trajectory. Due to the design of our setup, falling atoms are much more likely to be trapped, which is why we limit our analysis to these cases.

The dipole trap is loaded in the following way: whenever the FPGA-based detection and control system registers a single atom in the evanescent field of the resonator mode, it triggers the trap light to switch on rapidly. However, the trap loading succeeds in only $\eta_0 \approx 1$ \% of the cases. This can be attributed to the finite overlap of the trap volume with the resonator mode, the initial kinetic energy of the atom and the finite time delay between atom detection and switching on the trap.

\subsection{Experimental sequences with single trapped atoms}
\label{sec:Exp_sequ}

\begin{figure}[t]
	\centering
	\includegraphics[width=8.5cm]{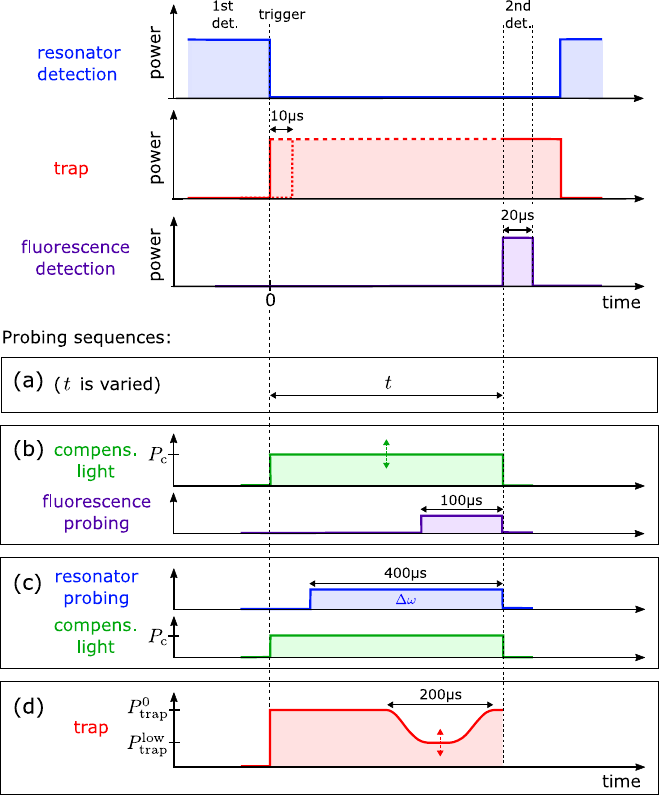}
	\caption{Basic experimental sequence and probing sequences for (a) the lifetime measurement, (b) the fluorescence measurement, (c) the measurement of the transmission spectra (see main manuscript) and (d) the measurement of the energy distribution of the trapped atoms (see section \ref{sec:Temp-meas}).}
	\label{fig:Exp-seq}
\end{figure}

Whenever the FPGA-based detection and control system registers a single atom in the evanescent field of the resonator mode ($1^\text{st}$ detection), it gives a trigger signal to start the experimental sequence for a measurement. The sequences used for the measurements presented in this work are illustrated in Fig.~S~\ref{fig:Exp-seq}. The basic sequence structure is common to all measurements and it is shown in the upper part of Fig.~S~\ref{fig:Exp-seq}. Just after a trigger signal the detection light is switched off and the trap light is switched on. Then follows a probing phase that is different for each measurement, see panels (a)-(d). At the end of the sequence, while the trap is still on, we apply a fluorescence detection pulse as described in the main part of the manuscript. This allows to verify whether there is an atom in the trap after the probing phase ($2^\text{nd}$ detection). For each measurement setting we also run a reference sequence, where the only difference is that the trap light switches on with a 10~$\mu$s-delay, such that a detected atom will not be trapped, as free-falling atoms leave the evanescent field within 1-2~$\mu$s. The reference sequence enables us to obtain an estimate of the time-dependent background signal originating from photons of the fluorescence laser beam scattered into our detector.

In the lifetime measurement no probe light is applied between $1^\text{st}$ and $2^\text{nd}$ detection, as depicted in panel (a) of Fig.~S~\ref{fig:Exp-seq}. We run four measurement sequences and four corresponding reference sequences for waiting times $t = \left\lbrace 0.1, 0.5, 1, 2 \right\rbrace $ ms, respectively. Those eight sequence runs are repeated a few thousand times to yield good measurement statistics. Figure~1(c) shows the background-corrected probability $\eta$ for finding a single atom at a waiting time, $t$, after switching on the trap. 

In the fluorescence measurement shown in Fig.~3(b) the compensation light switches on together with the trap light and switches off just before the $2^\text{nd}$ detection, as illustrated in panel (b) of Fig.~S~\ref{fig:Exp-seq}. During the last 100~$\mu$s before the $2^\text{nd}$ detection, the external probe light is sent onto the atom and we record photons outcoupled into the nanofiber. We run eight measurement sequences, in which we vary the compensation laser power, $P_\text{c} \approx \left\lbrace 0,\,275,\,365,\,455,\,545,\,635,\,725,\,815 \right\rbrace $~$\mu$W, and one reference sequence at the highest value of~$P_\text{c}$. The sequence duration is about 500~$\mu$s. 

The sequence structure for measuring the transmission spectra, shown in Fig.~4, is similar to the fluorescence measurement sequence. The only difference is that we send the probe light through the coupling fiber onto the resonator for a probing duration of 400 $\mu$s. We perform in total twenty-two measurement sequences, each for a different light-resonator detuning $\Delta \omega$, as depicted in panel (c) of Fig.~S~\ref{fig:Exp-seq}.

\subsection{Trap-induced resonator heating}
\label{sec:Heating}

When the trapping beam impinges on the bottle microresonator, a small fraction of the trap power is absorbed by the resonator, which causes a temperature increase of the latter. As a consequence, the refractive index of the resonator material (silica) and the resonator diameter change. Both effects lead to a shift of the resonator frequency, which is accompanied by a transmission increase of the detection light through the coupling fiber.

During the 80 ms-long detection window (see section \ref{sec:Exp_loading}), the resonator frequency is not stabilized. Whenever the trap light switches on upon an atom detection event for, e.g., 2~ms, we observe a temperature-induced shift of the resonator frequency of up to 2-3 MHz. The subsequent cooling down after switching the trap off has a larger time constant than the heating. Therefore, if multiple atom events occur successively, we observe a slow drift of the resonator frequency, which can increase the number of false detection events, where no atom is present.

To prevent this, we send an additional laser beam onto the resonator along the $-z^\prime$-direction, with the inverse on-off switching pattern as the trap light, such that it is always on when the trap light is off and vice versa. This heating compensation laser has a wavelength of 980~nm, a power of up to 100~mW and a focus diameter of $\approx~50$~$\mu$m. These parameters allow us to counteract the heating effect on the time scale of our experiment while exerting only negligible optical forces on the atoms prior to detection. 

\subsection{Energy distribution and resulting position distribution of trapped atoms}
\label{sec:Temperature}

\subsubsection{Measurement of the energy distribution}
\label{sec:Temp-meas}

In order to get an estimate of the temperature of the trapped atoms, we use a technique proposed and demonstrated in \cite{alt2003single}. The idea is to measure the fraction of atoms remaining in the trap, when the trap potential is adiabatically lowered from its original depth $U_0$ to a depth $U_\text{low}$, as a function of $U_\text{low}$. During this process the energy of the trapped atom also decreases from $E_0$ to $E$ (''adiabatic cooling''). A criterion for the adiabaticity of the trap depth variation, $U(t)$, is that the change of the atomic oscillation frequencies, $\omega(t)$, should be small at any time, $t$, during the variation: $\dot{\omega}(t) \ll \omega^2(t)$. 

\begin{figure}
	\centering
	\includegraphics[width=8cm]{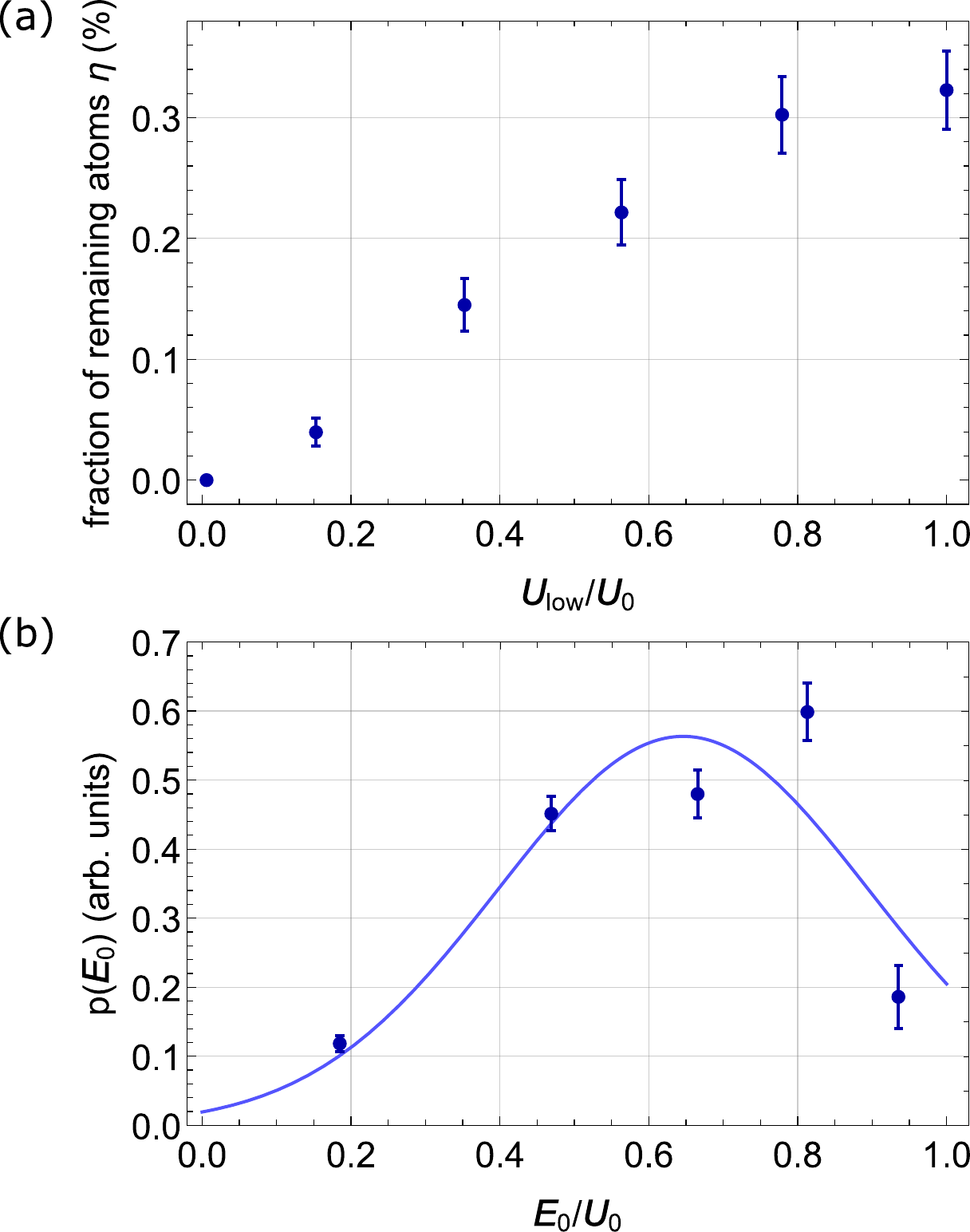}
	\caption{(a) Measured fraction of atoms remaining in the lowered trap potential as a function of its depth $U_\text{low}$. (b) Energy distribution: derivative of the data in (a) with rescaled abscissa, see text. The solid line is a Gaussian fit to the data.}
	\label{fig:Energy-distr}
\end{figure}

For the measurement we use the basic sequence as illustrated in Fig.~S~\ref{fig:Exp-seq}, but with the trap part replaced by the time-dependence shown in panel (d). When an atom is detected in the evanescent field of the resonator mode, the trap switches on with a power of $P_\text{trap}^0=20$ mW. Then we ramp down the trap power within 75~$\mu$s to a value $P_\text{trap}^\text{low}$ using an AOM and hold it for 50~$\mu$s. The hold time is longer than the longest oscillation periods along the transverse axes, $2 \pi/\omega_{y,z}$, thereby ensuring that an atom with a total energy $E > U_\text{low}$ indeed leaves the trap. Afterwards, we ramp up the trap power to the original value $P_\text{trap}^0$ with the time-reversed function, succeeded by a fluorescence detection pulse to test the presence of the atom. We repetitively run six measurement sequences with $P_\text{trap}^\text{low}/P_\text{trap}^0 =\left\lbrace 1,0.8,0.6,0.4,0.2,0\right\rbrace $, respectively, and a reference sequence without the trap power ramp ($P_\text{trap}^\text{low}/P_\text{trap}^0 =1$). Figure~S~\ref{fig:Energy-distr}(a) shows the measured fraction $\eta$ of atoms that are present after the trap lowering as a function of the trap depth $U_\text{low}$. The atoms that contribute to this signal thus have an energy $E \leq E_\text{max} \equiv U_\text{low}$ in the lowered trap. 

In order to connect the energy of the fraction of atoms that is lost when reducing the trap depth with the energy they possessed in the full potential, we make the assumption of adiabatic lowering of the trap. In this case, the atoms will always stay in well defined energy eigenstates. Comparing the eigenenergies for the different trap depths then allows us to calculate the initial energy of the lost atoms. From this, together with a pointwise derivative ($\Delta \eta /\Delta U_\text{low}$) of the measured data in Fig.~S~\ref{fig:Energy-distr}(a) we obtain an estimate of the energy distribution in the trap shown in Fig.~S~\ref{fig:Energy-distr}(b). As is apparent, the data cannot be described by a Boltzmann distribution, but is rather peaked towards higher energies. For our further analysis, we fit a Gaussian function to the data. We find its maximum at an energy $E_0 \approx 2/3\, U_0 = k_\text{B} \times 2$ mK. This agrees with what we expect from our trap loading mechanism.

\subsubsection{Position distribution}
\label{sec:Temp-pos-distr}

For modeling the theoretically expected signals in Fig.~3(b) and 4(a), we require the average position distribution of the atoms in the trap. For this purpose, we calculate classical atomic trajectories of a trapped atom according to the measured energy distribution, see section \ref{sec:Temp-meas}. For each possible value of the total energy $E_0$ we calculate $500$ classical trajectories: we pseudo-randomly generate a start position and a start velocity vector, under consideration of energy conservation, and numerically solve the equations of motion. We then histogram the occurring positions in bins of size $\Delta x \times \Delta y \times \Delta z = \left( 20 \times 100 \times 100 \right) $~$\text{nm}^3$ within a volume of 6.4~$\mu \text{m}^3$ around the trap center. To get the final position distribution, we average all histograms obtained for a fixed energy over the Gaussian energy distribution from section \ref{sec:Temp-meas}.

\subsection{Light shift compensation}
\label{sec:LSC}

\subsubsection{Compensation laser offset-lock}
\label{sec:LSC-absorb}

In order for the compensation laser to predominantly act on the excited state, $5P_{3/2}$, its frequency has to be close to resonance with the $5P_{3/2} \rightarrow 5D_{5/2}$ fine-structure transition. To ensure stable operation, we lock the laser to this transition with a frequency offset of 
\begin{equation}
\Delta_\text{c}/2 \pi = \nu_{5P_{3/2} \rightarrow 5D_{5/2}}-\nu_\text{c} \approx 927  \;\text{MHz}\,.
\label{eq:Locking_Delta-c}
\end{equation}
To do so, we perform absorption spectroscopy of the $5P_{3/2} \rightarrow 5D_{5/2}$ transition, with a two-step excitation scheme along the lines of \cite{kalatskiy2017frequency}. 

As the first step, a $780$~nm pump laser, which is locked to the $5S_{1/2} \rightarrow 5P_{3/2}$ transition and has an optical power of about $10$~mW, excites Rb atoms in a heated vapor cell ($\approx 50$~$^\circ$C). The second excitation step is performed with about 2 mW of a counter-propagating $776$~nm probe laser (New Focus, Velocity TLB 6700). To enable locking the laser with an offset to the atomic transition, we send the probe light through an electro-optical modulator (EOM) before it enters the vapor cell. Thus, the light contains three components: one at the carrier frequency $\nu_\text{carrier}$ and two sideband frequencies $\nu_\text{carrier} \pm \nu_\text{EOM}$, where $\nu_\text{EOM}$ is the RF-driving frequency for the EOM. The frequency of the Velocity laser is set such that the ''blue'' sideband ($\nu_\text{c} + \nu_\text{EOM}$) is shifted into resonance with the $5P_{3/2} \rightarrow 5D_{5/2}$ transition. Measuring the probe transmission through the vapor cell on a photodiode, we observe an absorption dip, which we use to generate an error signal via a Pound-Drever-Hall-type scheme. The error signal is fed back to the Velocity laser to stabilize the laser frequency. 

\subsubsection{Calculation of the detuning in Fig. 3(a)}
\label{sec:LSC-calc}

For the calculations of the light shifts induced by the trap and compensation laser fields we use the formalism given in \cite{le2013dynamical}, as it allows us to determine the energy level shifts of a multilevel atom interacting with a far-detuned light field, $\bm{E}$, of arbitrary polarization, $\bm{\epsilon}$. The interaction between the light field and the induced atomic dipole moment is described by the Stark operator 
\begin{equation}
V_\text{Stark}= - \bm{d} \cdot \bm{E} =- \alpha \bm{E}^2 \,,
\label{eq:Stark-op}
\end{equation}
where we used the expression for the operator of the electric dipole of the atom, $\bm{d}=\alpha \bm{E}$, in which $\alpha$ is the atomic polarizability. The polarizability of a certain fine-structure state $|nJ\rangle$ is proportional to the sum of the dipole matrix elements over the fine-structure states $|n^\prime J^\prime\rangle $, to which $|nJ\rangle$ can couple: 
\begin{align}
\alpha_{n,J}^{(K)} \approx \; & (-1)^{K+J+1}\sqrt{2K+1}\nonumber\\ 
& \times \sum_{n'J'}(-1)^{J'}\begin{Bmatrix}
1 & K & 1 \\
J & J' & J
\end{Bmatrix}|\left\langle n^\prime J^\prime||\bm{d}||nJ\right\rangle|^2\nonumber \\
& \times \frac{1}{\hbar}\text{Re}\bigg(\frac{1}{\omega_{n'J'nJ}-\omega} +\frac{(-1)^{K}}{\omega_{n'J'nJ}+\omega}\bigg) \,.
\label{eq:Fam_polarizability}
\end{align}
Here, $J$ is the quantum number for the total angular momentum $\bm{J}$ of the electron, and $n$ is the set of quantum numbers $\left\lbrace n, L, S, I\right\rbrace $. $L$ and $S$ are the quantum numbers for the total orbital angular momentum and the total spin of the electrons, respectively, and $I$ is the quantum number for the nuclear spin. $K=0,\,1,\,2$ in Eq. (\ref{eq:Fam_polarizability}) indicates the reduced dynamical scalar, vector and tensor polarizability of an atom in the fine-structure level $|nJ\rangle$, respectively. Furthermore, $\omega_{n'J'nJ}$ is the angular frequency of the $|nJ\rangle \rightarrow  |n^\prime J^\prime\rangle$ transition, and  $\omega$ denotes the angular frequency of the light field. 

In general, all three parts of the dynamical polarizability contribute to the Stark operator. In the hyperfine-structure (hfs) basis $\left\lbrace \ket{(nJ)FM}\right\rbrace $, its matrix elements $V_\text{Stark}^{FMF^\prime M^\prime}=\bra{(nJ)FM} V_\text{Stark} \ket{(nJ) F^\prime M^\prime}$ are given by
\begin{align}
V_\text{Stark}^{FMF^\prime M^\prime}=&\frac{|\mathcal{E}|^2}{4} \sum_{q=-K,...,K}^{K=0,1,2}\alpha_{n,J}^{(K)} \{\boldsymbol{u}^*\otimes \boldsymbol{u}\}_{Kq}\nonumber\\
&\times (-1)^{J+I+K+q-M}\sqrt{(2F+1)(2F'+1)}\nonumber\\
&\times \begin{pmatrix}
F & K & F' \\
M & q & -M'
\end{pmatrix}\begin{Bmatrix}
F & K & F' \\
J & I & J 
\end{Bmatrix}  \label{eq:Fam_Starkoperator}
\end{align}
Here, $F$ is the quantum number of the total angular momentum $\bm{F}=\bm{J}+\bm{I}$ of the atom and $M$ is the quantum number of the projection onto the quantization axis ($z$). $\mathcal{E}$ is the field amplitude and $\bm{u}$ is the polarization unit vector. The compound tensor components are defined as
\begin{align}
\{\boldsymbol{u}^*\otimes \boldsymbol{u}\}_{Kq}=&\sum_{\mu,\mu'=0,\pm1}(-1)^{q+\mu'}u_\mu u^*_{-\mu'}\nonumber\\ 
& \times \sqrt{2K+1} \begin{pmatrix}
1 & K & 1 \\
\mu & -q & \mu^\prime
\end{pmatrix} \label{eq:uu}
\end{align}
with the spherical tensor components of the polarization vector $\boldsymbol{u}$ in the Cartesian coordinate frame $\{x,y,z\}$: $u_{-1} = (u_x-iu_y)/\sqrt{2}$, $u_0 = u_z$, and $u_1 = -(u_x + iu_y)/\sqrt{2}$.\\

\begin{figure*}
	\centering
	\includegraphics[width=0.85\textwidth]{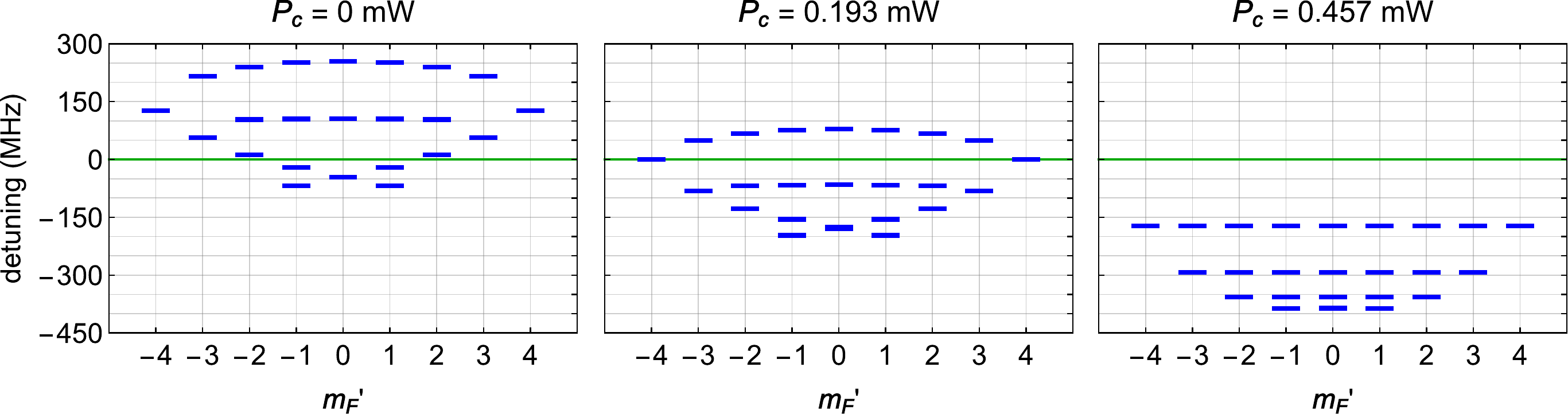}
	\caption{Detunings of the light shifted frequencies of the transitions between the $ 5S_{1/2},F=3 $ ground state and the $5P_{3/2}$ excited state manifold with respect to $\omega_0$, for $\left\lbrace P_\text{trap},\,w_\text{trap}=w_\text{c},\,\Delta_\text{trap}/2\pi,\,\Delta_\text{c}/2\pi\right\rbrace =\left\lbrace 18.7\,\text{mW},\,3.5\,\mu\text{m},\, 1.68\,\text{THz},\,927\,\text{MHz}\right\rbrace $ and three different values of the compensation laser power: a) in absence of the compensation laser, b) for the case where the light shift of the cycling transition is canceled, c) for the case of vanishing tensor light shifts. }
	\label{fig:LS-calc_F4manifold}
\end{figure*}

In order to find the new energy eigenvalues for a fine-structure state of the atom subjected to the trap and compensation light fields, we diagonalize the full interaction Hamiltonian
\begin{equation}
H_\text{int}=V_\text{Stark}^\text{trap}+V_\text{Stark}^\text{c}+V_\text{hfs} \,,
\label{eq:Hamil}
\end{equation} 
where $V_\text{hfs}$ is the operator of the atomic hfs interaction. It is diagonal in the hfs basis and its non-zero matrix elements are given by 
\begin{multline}
	\left\langle (n J) F M\left|V^{\mathrm{hfs}}\right| (n J) F M\right\rangle=\frac{1}{2} \hbar A_{\mathrm{hfs}} G \\
	+\hbar B_{\mathrm{hfs}} \frac{\frac{3}{2} G(G+1)-2 I(I+1) J(J+1)}{2 I(2 I-1) 2 J(2 J-1)}
\end{multline}
where $G=F(F+1)-I(I+1)-J(J+1)$.\\

As we are interested in the modification of the frequency of the $\left( 5S_{1/2},F=3 \right)  \rightarrow 5P_{3/2}$ transition due to the two light fields, we have to calculate the light shifts for the ground state, $|5,\frac{1}{2}\rangle$, and the excited state, $|5,\frac{3}{2}\rangle$. In the calculation of the polarizabilities $\alpha_{5,1/2}^{(K)}$ and $\alpha_{5,3/2}^{(K)}$, we take the coupling to many excited state levels $|n^\prime J^\prime\rangle$ into account. However, for our light parameters, the coupling to the transitions $5S_{1/2} \rightarrow 5P_{1/2},\,5P_{3/2}$, and $5P_{3/2} \rightarrow 5S_{1/2}$ as well as $5P_{3/2} \rightarrow 5D_{3/2},\,5D_{5/2}$ predominantly contributes to the light shift of the ground and excited state, respectively.
For the calculation shown in Fig.~3(a) we assumed linear polarization of the light fields, $\bm{\epsilon}=(0,0,1)$. In this case the contribution of the vector polarizability to the light shift is zero, $\alpha_{n,J}^{(K=1)}=0$. Furthermore, for $J=1/2$ the tensor polarizability vanishes. Thus, the ground state only experiences a scalar light shift, i.e. all Zeeman levels are shifted by the same amount. For the parameter values $\left\lbrace P_\text{trap},\,w_\text{trap}=w_\text{c},\,\Delta_\text{trap}/2\pi,\,\Delta_\text{c}/2\pi\right\rbrace =\left\lbrace 18.7\,\text{mW},\,3.5\,\mu\text{m},\, 1.68\,\text{THz},\,927\,\text{MHz}\right\rbrace $ we find the ground state light shift in the trap center, $\left(  x_0=205\;\text{nm},\,y_0=0,\,z_0=0 \right) $, to be
\begin{equation}
\delta_{\text{g},y} \approx -129 \,\text{MHz} \;\; \text{or} \;\; \delta_{\text{g},z^\prime} \approx -118 \,\text{MHz} \,,
\label{eq:GS-LS}
\end{equation}
depending on whether the trap light is polarized along the $y$-direction or along the $z^\prime$-direction. For this calculation, we assumed that the trap and compensation light fields are Gaussian beams with their waists lying on the resonator surface, such that their intensities at the trap center are given by
\begin{equation}
I_l=\frac{2 P_l}{\pi w_l} \times f_\text{refl}(x_0),
\label{eq:Int_trapcenter}
\end{equation}
where the index $l$ stands for trap or compensation light, and $f_\text{refl}(x_0)$ and is a factor taking into account the intensity modulation due to the partial standing wave pattern along the $x$-axis. It is $f_{\text{refl},y}(x_0)=1.43$ and $f_{\text{refl},z^\prime}(x_0)=1.31$ for the two different polarizations, respectively.

To obtain the change in the transition frequency, we also calculate the excited state light shift. To do this, we again diagonalize the Hamiltonian in Eq. (\ref{eq:Hamil}), which first yields the total shift for each $|F^\prime m_F^\prime \rangle$-level of the hyperfine manifold of the $5P_{3/2}$ state
\begin{equation}
\delta_\text{tot}^{F^\prime m_F^\prime}(P_\text{c})=\delta_\text{hfs}^{F^\prime}+\delta_\text{e}^{F^\prime m_F^\prime}(P_\text{c})\,,
\label{eq:ES-shift}
\end{equation}
consisting of the $F^\prime$-dependent hfs shift and the $F^\prime$- and $m_F^\prime$-dependent as well as $P_\text{c}$-dependent light shift. Finally, the detuning of each $|F m_F \rangle \rightarrow |F^\prime m_F^\prime \rangle$ transition with respect to the unperturbed frequency of the $\left(5 S_{1/2}, F=3\right) \rightarrow \left(5 P_{3/2}, F^\prime=4\right)$ transition is given by
\begin{equation}
\delta \omega(P_\text{c}) = \delta_\text{tot}^{F^\prime m_F^\prime}(P_\text{c})-\delta_\text{hfs}^{F^\prime=4}-\delta_\text{g} \,,
\label{eq:detuning}
\end{equation}
where the hyperfine shift of the $F^\prime=4$-level with respect to the $5 P_{3/2}$ fine structure level amounts to $\delta_\text{hfs}^{F^\prime=4}=100.205 \,\text{MHz}$ \cite{steck2010rubidium}, and $\delta_\text{g}$ is given in Eq. (\ref{eq:GS-LS}).

Figure~S~\ref{fig:LS-calc_F4manifold} shows the detuning for the excited state Zeeman manifold for our experimental values given above and three discrete values of the compensation laser power. The calculation is performed for linear polarization of the two light fields along the $y$-axis. If $P_\text{c}=0$~mW, the detuning is solely caused by the trap light. For $P_\text{c}=0.193$~mW both scalar and tensor contributions to the light shift are reduced. The cycling transitions to the outermost Zeeman levels are shifted back to the unperturbed transition frequency, which is the experimental setting we are aiming for. At a power of $P_\text{c}=0.457$~mW the tensor light shift vanishes, and the original excited state hyperfine structure is recovered, although now at a detuning of about 170~MHz from the unperturbed transition frequency.\\

From our fit to the data of the fluorescence measurement (see section \ref{sec:Models_LSC}) we conclude that there is a mismatch between the beam radii of the trap and compensation fields of $w_\text{c}/w_\text{trap} \approx~1.5$. In addition, the trap light field is not fully linearly polarized such that also the vector polarizability now contributes to the light shifts of the ground and excited state ($\alpha_{n,J}^{(K=1)} \neq 0 $). 

\begin{figure}
	\centering
	\includegraphics[width=8cm]{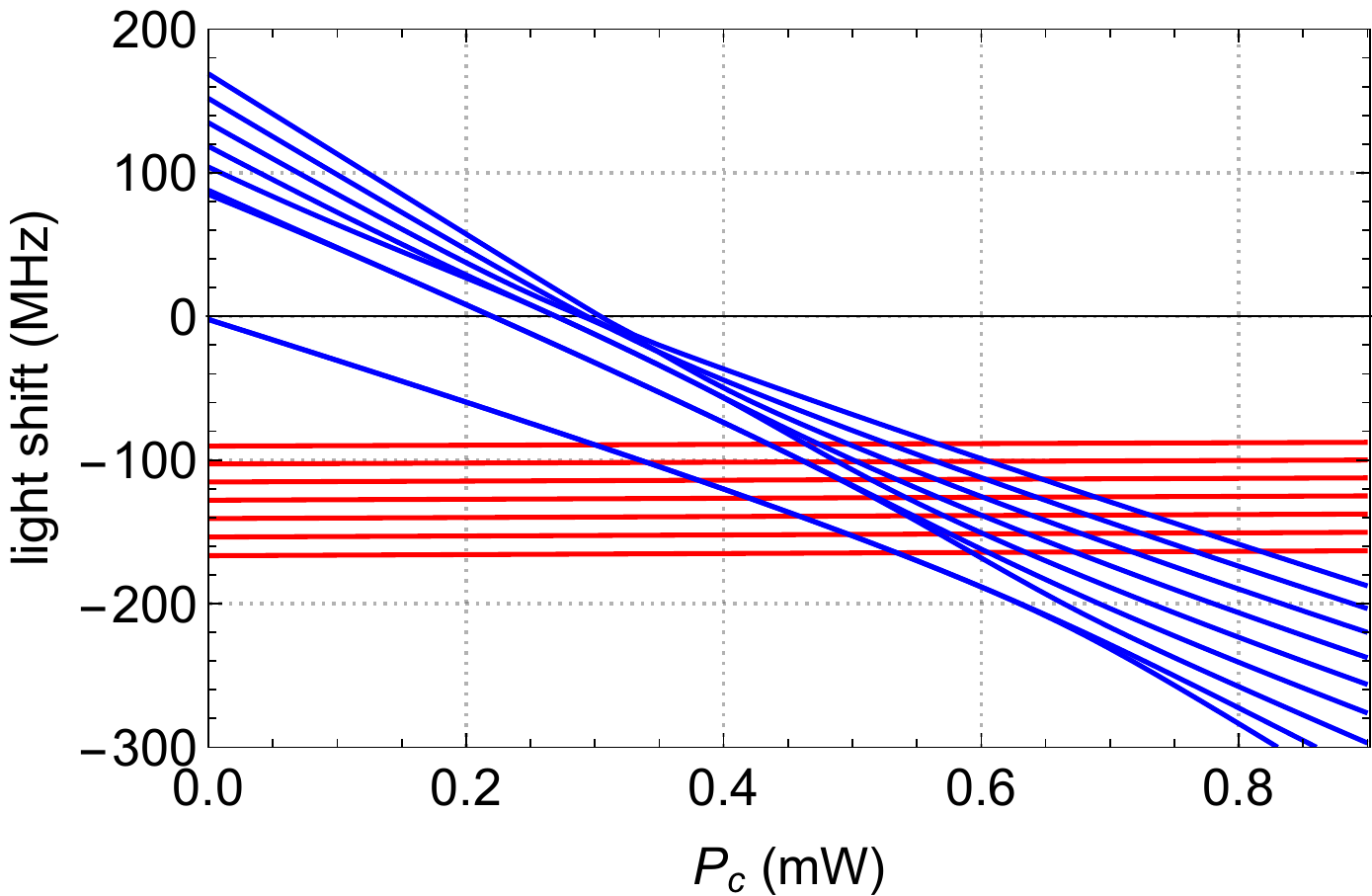}
	\caption{Calculation of the two-color light shifts of the $5S_{1/2},F=3$ ground state (red) and the $5P_{3/2},F^\prime=4$ excited state (blue), for an atom in the trap center, with two modifications compared to Fig.~3(a): a reduced overlap with linear polarization of the trap and compensation light fields ($\bm{\epsilon}=(u_x,u_y,u_z)=(0,0.98,0.20\,i)$, corresponding to about 96~\% overlap) and a mismatch of the beam radii at the position of the atom ($w_\text{c}/w_\text{trap} \approx 1.5$).}
	\label{fig:LS-calc_Ell02}
\end{figure}

Taking these effects into account, we again calculate the light shifts as a function of the compensation laser power for an atom in the trap center, see Fig.~S~\ref{fig:LS-calc_Ell02}. The light shift of the $ 5S_{1/2},F=3$ ground state ($5P_{3/2},F^\prime=4 $ excited state) Zeeman manifold is plotted in red (blue). Due to the vector contribution, the ground state light shift is no longer independent of the Zeeman-state. The $F=3 \rightarrow F^\prime=4$ transition frequency is shifted back to its unperturbed value in the power region around $P_\text{c} = 600$~$\mu$W, where the excited state light shifts cross the ground state light shifts. This power range is shifted to larger values compared to the equivalent situation in Fig.~3(a), which is indicated by the green shaded area. The reason is that the beam radius of the compensation laser at the position of the atom is now larger, and thus more compensation laser power is needed in order to reach the intensity required for cancelation of the trap-induced light shift.  

\subsection{Theoretical models/predictions}
\label{sec:models}
In the following, we describe how we obtain the theoretical predictions for the fluorescence measurement shown in Fig.~3(b) and for the compensated transmission spectrum plotted in Fig.~4(a). In the first case, the trapped atom is driven with an external resonant light field (''atom-drive''), which is sent through the trap optics. A certain fraction of the scattered photons is coupled into the resonator, and from the resonator into the coupling fiber with the rate $\kappa_\text{ext}$. The resonator output field is given by 
\begin{equation}
\begin{aligned}
\left\langle \hat{a}^\text{fluor}_{\text{out}}\right\rangle &=-i \sqrt{2 \kappa_\text{ext}}\left\langle \hat{a} \right\rangle \,,
\label{eq:aout_LSC}
\end{aligned}
\end{equation}
where $\left\langle \hat{a} \right\rangle$ is the expectation value of the photon annihilation operator for the resonator field. 

In the transmission measurement, the probe field, $\left\langle \hat{a}_\text{in} \right\rangle$, is sent into the coupling fiber (''resonator-drive''), and we are interested in the field transmitted through the coupling fiber, which is given by
\begin{equation}
\begin{aligned}
\left\langle \hat{a}^\text{trans}_{\text{out}}\right\rangle &=\left\langle \hat{a}_\text{in} \right\rangle-i \sqrt{2 \kappa_\text{ext}}\left\langle \hat{a} \right\rangle \, .
\label{eq:aout_Rabi}
\end{aligned}
\end{equation}
The resonator field $\left\langle \hat{a} \right\rangle$ in both the Eqs. (\ref{eq:aout_LSC}) and (\ref{eq:aout_Rabi}) is obtained by solving the master equation of the atom-resonator system 
\begin{equation}
\frac{d \hat{\rho}}{dt} = -\frac{i}{\hbar} [\hat{H},\hat{\rho}] + \mathcal{L} \hat{\rho} \;, 
\label{eq:MasterEq}
\end{equation}
where $\hat H$ is the Hamiltonian of the system and $\mathcal{L}$ is the Lindblad superoperator \cite{Carmichael1999}. For the simulations of both measurements we assume a two-level atom and use the Jaynes-Cummings Hamiltonian in the rotating wave approximation to describe the atom-resonator field interaction
\begin{equation}
\frac{\hat{H}}{\hbar}=\Delta_\text{rl}\hat{a}^\dagger \hat{a}+\Delta_\text{al}\hat{\sigma}_{+}\hat{\sigma}_{-} +g(\hat{a}^\dagger \hat{\sigma}_{-}+\hat{a} \hat{\sigma}_+) + \frac{\hat{H}_\text{drive}}{\hbar}\,,
\label{eq:Ham2Level}
\end{equation}
where $\Delta_\text{rl}=\omega_\text{r}-\omega_\text{probe}$ ($\Delta_\text{al}=\omega_\text{a}-\omega_\text{probe}$)~is~the resonator-light (atom-light) detuning, $\hat{\sigma}_+$ ($\hat{\sigma}_-$) the atomic excitation (deexcitation) operator, and $g$ the atom-light coupling strength. The last term $\hat{H}_\text{drive}$ describes the coupling to the driving field. The Lindblad superoperator is given by
\begin{equation}
\begin{aligned}
\mathcal{L} =\kappa_\text{\rm tot}& ( 2\hat a\hat{\rho} \hat a^\dagger-\hat a^\dagger  \hat a \hat{\rho}-\hat{\rho}  \hat a^\dagger \hat a) \\
+ \gamma &( 2\hat \sigma_-\hat{\rho} \hat \sigma_+ -\hat \sigma_+ \hat \sigma_- \hat{\rho}-\hat{\rho}  \hat \sigma_+ \hat \sigma_-) 
\label{eq:Lind2Level}
\end{aligned}
\end{equation}
where $\kappa_\text{\rm tot}=\kappa_0+\kappa_\text{\rm ext}$ with $\kappa_0$ being the intrinsic decay rate of the resonator. 

\subsubsection{Fluorescence as a function of compensation laser power}
\label{sec:Models_LSC}

In the fluorescence measurement we used the ''atom-drive'' setting, which is described by the driving Hamiltonian
\begin{equation}
\hat{H}_\text{drive}^\text{fluor}=i \hbar \frac{\Omega}{2} (\hat \sigma_--\hat \sigma_+)\,,
\label{eq:Hdrive_LSC}
\end{equation}
where $\Omega$ is the Rabi frequency of the external probe laser. In the case of weak driving, one can analytically solve the master equation in steady-state ($d \hat{\rho}/dt =0$) and the outcoupled power yields in first approximation
\begin{equation}
\begin{aligned}
\left|\langle \hat{a}^\text{fluor}_\text{out}\rangle\right|^2  = & 2 \kappa_\text{ext} \\ 
&\times \left|\frac{g \Omega/2}{g^2+\left( \gamma +i \Delta_{\text{al}}\right) \left( \kappa_{\text{\rm tot}}+i \Delta_{\text{\rm rl}}\right)}\right|^2 \,.
\end{aligned}
\label{eq:powerout_LSC}
\end{equation}
The probe light was resonant to the unperturbed atomic transition and 10 MHz detuned to the resonator frequency.

In order to fit the measured fluorescence in Fig.~3(b), we take two effects into account that influenced the measurement. First, the shift of the maximum of the fluorescence spectrum to a higher compensation power, $P_\text{c}^\text{max} \approx 500$ $\mu$W, suggests that the compensation laser intensity at the position of the atom was smaller than expected. We attribute this to a difference in the beam radii of the trap and compensation lasers, $w_\text{c} > w_\text{trap}$, at the position of the atom, caused by chromatic aberrations of the trap optics. Thus, more compensation laser power is needed to reach the intensity required to cancel the trap-induced light shift.

Another consequence of this foci displacement is that the trap and compensation fields are not perfectly spatially mode-matched, in spite of being coupled into the same single-mode fiber before entering the trap optics. Consequently, the two-color light shift of the trapped atom, given by Eq.~(\ref{eq:detuning}), and, thus, the atom-light detuning become dependent on the position, $\bm{r}=(x,y,z)$, of the atom in the trap: $\Delta_\text{al}(P_\text{c},\bm{r})$=$\delta \omega (P_\text{c},\bm{r})$. This, in conjunction with the atomic motion, results in a broadening of the measured fluorescence spectrum. Note, that the mismatch of the two intensity patterns manifests itself mainly in the transverse ($yz$-) plane, while it is very limited over the short extent of the trapping potential in $x$-direction of $\approx 300$ nm.

The broadening mechanism just described does, however, not fully explain the width (FWHM) of the measured fluorescence spectrum of $\approx 300$ $\mu$W. Therefore, as the second effect, we take into account that the polarization vector of the trap- and compensation fields could be subject to residual elliptical polarization components, potentially arising from stress-induced birefringence of the large vacuum viewport. This leads to an additional vector light shift, as discussed in section \ref{sec:LSC-calc} (Fig.~S~\ref{fig:LS-calc_Ell02}). 

For a certain position $\bm{r}$, we calculate the light shifts as shown for the trap center position in Fig.~S~\ref{fig:LS-calc_Ell02}. Using the position-dependent light shifts, we average the fluorescence spectrum, calculated according to Eq. (\ref{eq:powerout_LSC}), over the detunings of all possible transitions between the $\left( 5 S_{1/2}, F=3\right) $- and $\left( 5 P_{3/2}, F^\prime=4\right) $-Zeeman manifolds, with respect to the unperturbed transition frequency. Here, we also take the position-dependency of the Rabi frequency $\Omega(\bm{r})$ and the coupling strength $g(\bm{r})$ into account. The latter is determined by the intensity distribution of the resonator mode, see section \ref{sec:WGM}.  
Finally, the calculated spectra are averaged over the position distribution of the atoms in the trap, which we derive from the measured energy distribution, see section \ref{sec:Temperature}. 

We fit this averaged fluorescence spectrum to the measured data with the ratio of the beam radii $w_\text{c}/w_\text{trap}$ and the ellipticity of the polarization vector as free parameters and obtain the best-fitting result for $w_\text{c}/w_\text{trap} \approx 1.5$ and a reduced overlap with linear polarization of $\sim$ 96~\%. 

\subsubsection{Transmission spectrum with light shift compensation}
\label{sec:Models_Rabi}

\begin{figure}[t]
	\centering
	\includegraphics[width=7.5cm]{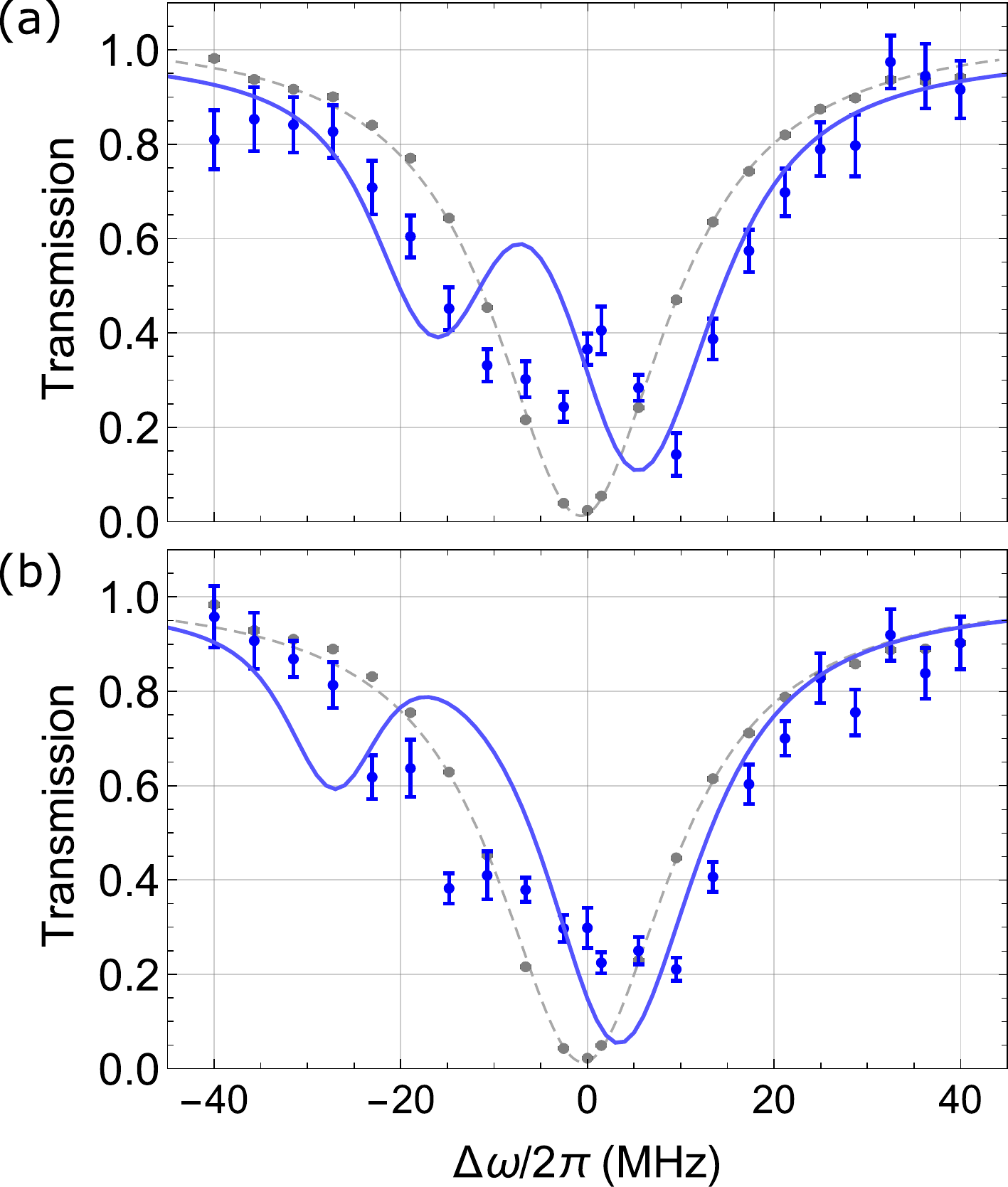}
	\caption{Normalized transmission spectra for compensation laser powers of (a) $P_\text{c} = 370\pm35$~$\mu$W and (b) $P_\text{c} = 410\pm40$~$\mu$W. The solid lines are theoretical predictions, see text.}
	\label{fig:Rabi_15p6mV}
\end{figure}

In the measurements of the transmission spectra we used the ''resonator-drive'' setting, which is described by the driving Hamiltonian
\begin{equation}
\hat{H}_\text{drive}^\text{trans}=i \hbar \sqrt{2 \kappa_\text{ext}} (\hat a -\hat a^\dagger)\,.
\label{eq:Hdrive_Rabi}
\end{equation}
Under the assumption of weak driving, see section \ref{sec:Models_LSC}, we obtain for the transmission through the coupling fiber
\begin{equation}
\begin{aligned}
T & =  \left|\frac{\left\langle \hat{a}^\text{trans}_{\text{out}}\right\rangle}{\left\langle \hat{a}_\text{in}\right\rangle}\right|^2 \\ & = \left| \frac{g^2 + \left( \gamma +i \Delta_\text{al}\right) \left( \kappa_0-\kappa_\text{\rm ext} +i \Delta_\text{rl}\right)}{g^2 + \left( \gamma +i \Delta_\text{al}\right) \left( \kappa_0+\kappa_\text{\rm ext} +i \Delta_\text{rl}\right)} \right|^2\,,
\label{eq:powerout_Rabi}
\end{aligned}
\end{equation}
where $\Delta_\text{rl} \equiv \Delta \omega$ is the resonator-light detuning plotted on the abscissa in Fig.~4. Here too, we have to consider the position-dependence of the two-color light shift originating from the mismatch of the beam radii $w_\text{c}/w_\text{trap}~\approx~1.5$, which we determined in section \ref{sec:Models_LSC}. Thus, the atom-light detuning is now given by $\Delta_\text{al}(P_\text{c},\bm{r})=\Delta \omega+\delta \omega(P_\text{c},\bm{r})$. 

For simplicity, we assume the trap- and compensation fields to be linearly polarized along the $z^\prime$-axis. Accordingly, we only take the coupling to the $F=3,m_F=\pm 3 \rightarrow F^\prime=4,m_F=\pm4$ cycling transition into consideration, see main part of the manuscript. 

In order to simulate a transmission spectrum for a certain compensation laser power, $P^\text{th}_\text{c}$, we calculate transmission spectra according to Eq. (\ref{eq:powerout_Rabi}) for each position $\bm{r}$, where again the spatially dependent coupling strength, $g(\bm{r})$, and detuning, $\delta \omega(\bm{r}; P^\text{th}_\text{c})$, enter. Then, we average the spectra for all considered positions over the position distribution as done for the fluorescence spectrum. We repeated this procedure for different values of $P^\text{th}_\text{c}$ in order to find the symmetric transmission spectrum plotted in Fig.~4(a).

\subsubsection{Transmission spectra for different compensation powers}
\label{sec:Models_Rabi_otherPc}

In addition to the transmission spectrum shown in Fig.~4(a), we measured two more spectra for the compensation laser powers $P_\text{c} = 370 \pm 35 $~$\mu$W and $P_\text{c} = 410 \pm 40$~$\mu$W, respectively. The spectra are shown in Fig.~S~\ref{fig:Rabi_15p6mV}. The solid lines are theoretical predictions, which were generated as described in section \ref{sec:Models_Rabi} for compensation laser powers of $P^\text{th}_\text{c}=440$~$\mu$W and $P^\text{th}_\text{c} = 490$~$\mu$W, respectively, which were chosen to have a similar ratio of $P^\text{th}_\text{c}/P_\text{c}$ as in the main part of the manuscript. Note that these spectra were measured at higher compensation laser powers, i.e. in a situation where the transitions to the other Zeeman levels of the $F^\prime=4$ manifold are close to the unperturbed transition frequency, while the cycling transition is already detuned, see Fig.~3(a). Thus the two-level atom approximation in our theory model does not apply anymore, and, as a consequence, the agreement between the measured data and the theoretical prediction is not as good as for the compensated spectrum in Fig.~4(a).

\subsubsection{Characteristic parameters of the investigated WGM}
\label{sec:WGM}
In the experiment, we couple the atoms to our bottle microresonator which has a central radius of $18.0$~$\mu$m and an axial curvature of about $0.014$~$\mu\text{m}^{-1}$. The TM polarized (transverse magnetic) axial mode with (estimated) modenumber $q=3$ is resonant with the atomic transition $F=3\rightarrow F'=4$. From our geometry 
we calculate \cite{louyer2005tunable}, the axial and radial intensity profile of the resonator mode and in particular the axial and radial dependency of the atom-resonator coupling strength $g$ which is used in our numerical models. 
In particular, we obtain from our geometry an axial spatial extension of our mode (caustic to caustic) of about $15$~$\mu$m and an atom-resonator coupling strength of $g\approx 2\pi\times 43.7$~MHz for the cycling transition $F=3,m_F=3\rightarrow F'=4, m_{F'}=4$, for an atom located at the caustic and at the surface of the resonator.

\bibliography{citation-suppl}